\documentclass[pdflatex,sn-mathphys-num]{sn-jnl}
\usepackage[defaultcolor=blue]{changes}
\usepackage{graphicx}%
\usepackage{multirow}%
\usepackage{amsmath,amssymb,amsfonts}%
\usepackage{amsthm}%
\usepackage{mathrsfs}%
\usepackage[title]{appendix}%
\usepackage{xcolor}%
\usepackage{textcomp}%
\usepackage{manyfoot}%
\usepackage{booktabs}%
\usepackage{algorithm}%
\usepackage{algorithmicx}%
\usepackage{algpseudocode}%
\usepackage{listings}%
\usepackage{subcaption}
\usepackage{siunitx}
\usepackage{hyperref}
\usepackage{array}
\usepackage{threeparttable}

\theoremstyle{thmstyleone}%

\theoremstyle{thmstyletwo}%

\theoremstyle{thmstylethree}%

\begin{document}

\title[Article Title]{DBMIF: A Deep Balanced Multimodal Iterative Fusion Framework for Air‑ and Bone‑Conduction Speech Enhancement}

\author[1,2]{\fnm{Yilei} \sur{Wu}}\equalcont{These authors contributed equally to this work.}\email{wyl21w@126.com}
\author[1,2,3]{\fnm{Changyan} \sur{Zheng}}\equalcont{These authors contributed equally to this work.}\email{echoaimaomao@163.com}

\author[1,2]{\fnm{Xingyu} \sur{Zhang}}\email{zhangxingyu1994@126.com}
\author[1,2]{\fnm{Yakun} \sur{Zhang}}\email{ykzhang1222@126.com}
\author[4]{\fnm{Chengshi} \sur{Zheng}}\email{cszheng@mail.ioa.ac.cn}
\author[5]{\fnm{Shuang} \sur{Yang}}\email{shuang.yang@ict.ac.cn}
\author*[1,2]{\fnm{Ye} \sur{Yan}}\email{yy\_taiic@163.com}
\author[1,2]{\fnm{Erwei} \sur{Yin}}\email{yinerwei1985@gmail.com}

\affil[1]{\orgname{Defense Innovation Institute, Academy of Military Sciences (AMS)},%
  \orgaddress{\city{Beijing}, \postcode{100071}, \country{China}}}

\affil[2]{\orgname{Intelligent Game and Decision Laboratory, Academy of Military Sciences (AMS)},%
  \orgaddress{\city{Beijing}, \postcode{100071}, \country{China}}}

\affil[3]{\orgname{High-tech Institute},%
  \orgaddress{\street{Fan Gong-ting South Street on the 12th}, \city{Weifang}, \postcode{261000}, \country{China}}}

\affil[4]{\orgname{Institute of Acoustics, Chinese Academy of Sciences},%
  \orgaddress{\city{Beijing}, \postcode{100190}, \country{China}}}

\affil[5]{\orgname{Institute of Computing Technology, Chinese Academy of Sciences},%
  \orgaddress{\city{Beijing}, \postcode{100190}, \country{China}}}

\abstract{The performance of conventional speech enhancement systems degrades sharply in extremely low signal-to-noise ratio (SNR) environments where air-conduction (AC) microphones are overwhelmed by ambient noise. Although bone-conduction (BC) sensors offer complementary, noise-tolerant information, existing fusion approaches struggle to maintain consistent performance across a wide range of SNR conditions. To address this limitation, we propose the Deep Balanced Multimodal Iterative Fusion Framework (DBMIF), a three-branch architecture designed to reconstruct high-fidelity speech through rigorous cross-modal interaction. Specifically, grounded in a multi-scale interactive encoder–decoder backbone, the framework orchestrates an iterative attention module and a cross-branch gated module to facilitate adaptive weighting and bidirectional exchange. To complement this dynamic interaction, a balanced-interaction bottleneck is further integrated to learn a compact, stable fused representation. Extensive experiments demonstrate that DBMIF achieves competitive performance compared with recent unimodal and multimodal baselines in both speech quality and intelligibility across diverse noise types. In downstream ASR tasks, the proposed method reduces the character error rate by at least 2.5\% compared to competing approaches. These results confirm that DBMIF effectively harnesses the robustness of BC speech while preserving the naturalness of AC speech, ensuring reliability in real-world scenarios. The source code is publicly available at \url{https://github.com/wyl516w/dbmif.}}




\maketitle

\section{Introduction}
\label{sec:intro}

Speech enhancement aims to recover intelligible and natural speech from noisy recordings and is a key front-end for remote communication, wearable hearing assistance, and human--computer interaction~\cite{wang2018supervised,healy2023progress,zhou2025ave}. With deep learning, single-channel systems have made substantial progress in handling non-stationary noise and generalizing across speakers, as reflected in recent benchmark iterations such as the Deep Noise Suppression (DNS) Challenge~\cite{dubey2024icassp}. However, real-world acoustic scenes often exhibit extremely low SNR conditions and highly non-stationary noise. Under these conditions, approaches relying solely on AC microphones often retain residual noise and introduce speech distortion because the sensor is directly exposed to environmental interference~\cite{zhou2020real}.

Bone-conduction (BC) sensors capture speech via skull vibration and naturally suppress airborne noise, providing strong robustness in adverse environments~\cite{huang2017wearable}. Yet BC speech typically exhibits pronounced high-frequency roll-off above approximately $\sim$1~kHz, limiting naturalness and intelligibility relative to AC speech~\cite{ito2010bone,nakajima2003non}. Beyond its limited bandwidth, BC speech is further affected by the propagation path and the transducer–skin contact, resulting in frequency-dependent spectral coloration and temporal misalignment that cannot be explained by a simple gain adjustment. These characteristics make AC and BC speech fusion non-trivial: naively mixing the two modalities may yield miscalibrated or imbalanced features, especially for high-frequency cues that are critical to consonant clarity. Prior studies have attempted to compensate for missing bands and path-induced distortions, from spectral equalization~\cite{kondo2006equalization} and bandwidth extension to recent Generative Adversarial Networks (GAN) based~\cite{pan2022cyclegan,hauret2023eben} restoration that recovers richer high-frequency structure. Despite these advances, the combined effects of bandwidth loss and modality-specific distortions still constrain BC speech when high perceptual quality is required, motivating the use of learned, adaptive fusion mechanisms that can calibrate cross-modal representations.

To jointly exploit the noise robustness of BC speech and the high-frequency detail of AC speech, multimodal fusion has attracted growing interest~\cite{hao2025lipgen}. Hardware systems have been developed that integrate accelerometers with BC microphones to enable robust voice activity detection and provide an auxiliary speech reference for enhancement~\cite{dusan2016system,lee2018bone}. Architecturally, separate modeling followed by fusion outperforms early concatenation in time-domain fully convolutional networks~\cite{yu2020time}. In the complex spectral domain, attention mechanisms with semi-supervised training can improve low-SNR performance~\cite{wang2022fusing}. Lightweight designs, such as DenGCAN, embed attention into skip connections to balance effectiveness and computational cost~\cite{kuang2024lightweight}. Nonetheless, existing multimodal methods remain fragile under the ultra-low SNR conditions typical of real deployments and struggle to maintain adaptability across diverse noise types. For example, during training at low SNR conditions, the AC speech quality is poor, whereas the BC channel is comparatively more noise-robust, therefore the BC speech branch tends to update more readily. As a result, BC modality that offers the largest immediate loss reduction dominates the gradients and suppresses the other modality. This drives the representations to collapse to a single modality and ultimately degrades overall performance.

These issues call for a framework that maintains sustained and balanced exchange between modalities. Cross-modal interaction enables the modalities to repeatedly refine one another, thereby preserving and exploiting their complementarity. In contrast, single-stage fusion typically combines the modalities at a sole integration junction and then processes only the fused stream, leaving no mechanism for continued cross-modal exchange. Compared with this single-stage strategy, iterative interaction can better limit dominance by any single modality and promote the learning of informative, noise-robust features.

Accordingly, to address the issues above, we propose a Deep Balanced Multimodal Iterative Fusion Framework (DBMIF). Unlike conventional static fusion approaches, DBMIF establishes a sustained cross-modal interaction mechanism that integrates multi-scale encoding with equilibrium-driven rebalancing. This design enables the system to dynamically adapt to instantaneous modality reliability, ensuring that the complementary strengths of AC and BC speech are fully utilized without introducing redundant parameters.

The main contributions of this work are summarized as follows:

\begin{enumerate}

\item A three-branch, multi-scale interactive encoder–decoder architecture is presented for robust AC and BC speech fusion. Through this framework, cross-scale feature consistency is preserved, thereby facilitating the effective exploitation of complementarity between the two modalities under real-world noise.

\item An early-stage deep iterative attention fusion mechanism is developed to adapt to instantaneous modality reliability. Specifically, higher attention is assigned to BC cues under harsh noise, while attention is progressively shifted to AC high-frequency cues as conditions improve.

\item A parameter-shared balanced-interaction bottleneck is introduced to recursively recalibrate fused representations. Consequently, stability is improved across a wide range of SNRs, and modal dominance is mitigated without increasing the parameter count.

\item Extensive experiments are conducted on public datasets. The proposed method is demonstrated to yield consistent gains over strong baselines in speech quality and intelligibility, alongside a significant reduction in the character error rate (CER) for downstream automatic speech recognition (ASR).
\end{enumerate}

\section{Related Work}
\label{sec:rel}

\subsection{AC Speech Enhancement}

Deep learning-based speech enhancement is generally categorized into time-frequency (T-F) domain and time-domain approaches~\cite{wang2018supervised}. T-F domain methods typically leverage the short-time Fourier transform to estimate spectral masks or directly regress clean spectra. Representative architectures include CRN~\cite{tan2018convolutional} and its complex-valued variant DCCRN~\cite{hu2020dccrn}, with the latter jointly modeling both magnitude and phase to mitigate phase mismatch artifacts. To further capture long-range temporal dependencies, advanced models like GaGNet employ attention mechanisms to integrate global context modeling with local spectral refinement~\cite{li2022glance}. Alternatively, time-domain models such as SEGAN~\cite{pascual2017segan} and Demucs~\cite{defossez2020real} operate on raw waveforms. While this facilitates implicit phase reconstruction, it often comes with significant computational costs. More recently, approaches based on self-supervised learning (SSL), such as HuBERT, have been introduced to enhance generalization by regularizing representations within pre-trained feature spaces~\cite{sato2025generic}. Despite these advancements, unimodal enhancement relying solely on AC signals remains inherently limited under extremely low SNR conditions, where the target speech is heavily masked by non-stationary noise.

\subsection{BC Speech Enhancement}

BC speech is characterized by pronounced high-frequency attenuation and nonlinear distortions resulting from propagation through the skull and soft tissue, rendering its restoration a challenging task. To mitigate these degradations, early studies explored linear approaches such as equalization~\cite{kondo2006equalization} and linear-phase reconstruction~\cite{shimamura2005reconstruction}. However, these methods struggled to model complex spectral mappings. Consequently, statistical mapping approaches based on Gaussian mixture models were introduced to improve intelligibility, yet they often suffered from over-smoothed spectra~\cite{toda2012statistical,shin2012survey}. With the advent of deep learning, strong modeling capabilities were leveraged through denoising autoencoders to simultaneously address noise suppression and bandwidth extension~\cite{liu2018bone,erzin2009improving}. Building on this, BLSTM-based models incorporating attention mechanisms were developed to better exploit temporal dependencies, achieving finer recovery of spectral details~\cite{zheng2018novel,zheng2019spectra}. Nevertheless, regression-based deep models tend to yield averaged spectra. To overcome this limitation, GAN-based schemes were investigated to reconstruct realistic high-frequency structures~\cite{pan2020bone,cheng2023speaker}. More recently, to ease the optimization difficulty of end-to-end mapping, sequential two-stage pipelines have been proposed to decouple the objectives of denoising and bandwidth extension~\cite{li2023two}.
Despite these advances, the intrinsic bandwidth limitation of BC signals and transmission-induced distortions continue to constrain restoration quality when high perceptual fidelity and naturalness are required.

\subsection{Multimodal Feature Fusion}

AC speech contains rich high-frequency details but is susceptible to noise, whereas BC speech offers robustness against ambient interference but suffers from limited bandwidth.
Therefore, multimodal fusion aims to integrate the strengths of both modalities.
Early studies employed input-level fusion by concatenating AC and BC features, reporting measurable performance gains under severe noise~\cite{dekens2013body}.
Subsequent approaches adopted late-fusion architectures that process each modality in a separate branch before merging, enabling wearable systems to jointly perform BC bandwidth compensation and denoising~\cite{huang2017wearable}.
Yu et al.~\cite{yu2020time} showed that independent modeling of both signals prior to fusion significantly outperforms simple input concatenation in time-domain fully convolutional networks.
In the complex spectral domain, Wang et al.~\cite{wang2022fusing} introduced an attention-based fusion module that adaptively weights AC and BC contributions according to noise levels, yielding substantial intelligibility gains at low SNR conditions.
More recently, lightweight designs such as DenGCAN incorporated attention mechanisms into skip connections to support real-time mobile deployment while maintaining competitive perceptual quality~\cite{kuang2024lightweight}.
Concurrently, multimodal enhancement approaches leveraging self-supervised learning (SSL) have attracted increasing attention.
For instance, audio-visual speech enhancement leverages SSL-pretrained encoders to reduce reliance on large paired datasets~\cite{lai2025ssl_avse}.
Similarly, recent AC-BC speech fusion works emphasize modality-specific enhancement coupled with noise-adaptive fusion, aligning with the trend toward reliability-aware integration~\cite{kim2025modality}.

Despite these advances, performance often degrades in extremely low SNR regimes common in real-world deployments.
Specifically, existing methods typically lack fine-grained reliability-aware fusion and stable cross-modal interaction.
Furthermore, balanced representation learning is often overlooked, leading to one modality dominating the optimization.

\section{Method}
\label{sec:meth}

\subsection{Problem Formulation}

Let $\mathbf{x}_{a} \in \mathbb{R}^{N}$ and $\mathbf{x}_{b} \in \mathbb{R}^{N}$ denote the noisy AC speech and the corresponding BC speech signals, respectively, where $N$ denotes the signal length.
Let $\mathbf{y} \in \mathbb{R}^{N}$ denote clean target speech waveform.
Direct time-domain processing typically treats the full-band signal uniformly, often ignoring the frequency-dependent characteristics of speech and noise.
To address this, we employ a pseudo-quadrature mirror filter (PQMF) bank~\cite{nguyen2002near} to decompose the input signals into subbands.

Given a discrete-time input modality $c \in \{a, b\}$, the $m$-th subband signal $\mathbf{s}_{c,m}$ is obtained by convolving signal $\mathbf{x}_c$ with the analysis filter $h_{m}$:
\begin{equation}
\label{eq:subband-def}
\mathbf{s}_{c,m}[n] = (\mathbf{x}_c * h_{m})[n], \quad m \in \{0,\dots,M-1\},
\end{equation}
where $*$ denotes the convolution operation.
The analysis filter $h_m$ and the corresponding synthesis filter $g_m$ are derived from a prototype low-pass filter $p[n]$ of length $L$, designed using a Kaiser-windowed sinc function:
\begin{equation}
\label{eq:analysis}
\begin{aligned}
h_{m}[n] &= 2p[L-1-n] \cos\!\left[\frac{(2m+1)\pi}{2M}\left(n-\frac{L-1}{2}\right) + (-1)^{m} \frac{\pi}{4}\right], \\
g_{m}[n] &= 2M p[n] \cos\!\left[\frac{(2m+1)\pi}{2M}\left(n-\frac{L-1}{2}\right) - (-1)^{m} \frac{\pi}{4}\right],
\end{aligned}
\end{equation}
where $0 \leq n \le L-1$ is the tap index, and $M$ is the number of subbands.

Based on the decomposed subband features, we propose a deep neural network, denoted as a parameterized mapping function $f_{\boldsymbol{\theta}}$, to estimate the clean subband components.
Here, $\boldsymbol{\theta}$ represents the set of learnable parameters.
The network takes the concatenated subbands of both modalities as input to exploit full-band contextual information.
Let $\mathbf{S}_c = [\mathbf{s}_{c,0}, \dots, \mathbf{s}_{c,M-1}]$ denote the collection of subbands for modality $c$.
The set of estimated clean subbands $\hat{\mathbf{S}}$ is obtained by:
\begin{equation}
\label{eq:inference}
\hat{\mathbf{S}} = f_{\boldsymbol{\theta}}(\mathbf{S}_a, \mathbf{S}_b).
\end{equation}
The enhanced time-domain waveform $\hat{\mathbf{y}}$ is then reconstructed via PQMF synthesis:
\begin{equation}
\label{eq:recon}
\hat{\mathbf{y}}[n] = \sum_{m=0}^{M-1} (\hat{\mathbf{s}}_{m} * g_{m})[n],
\end{equation}
where $\hat{\mathbf{s}}_{m}$ is the $m$-th estimated subband from $\hat{\mathbf{S}}$.
Finally, the network is trained by minimizing the loss function $\mathcal{L}$ between the estimated speech and the ground truth:
\begin{equation}
  \min_{\boldsymbol{\theta}} \, \mathcal{L}(\mathbf{y}, \hat{\mathbf{y}}).
\end{equation}

In our experiments, we decompose the input signals into $M=4$ subbands using the PQMF bank with a prototype filter of length $L=64$.

\subsection{Model Overview}

\begin{figure}[ht]
  \centering
  \includegraphics[width=\textwidth]{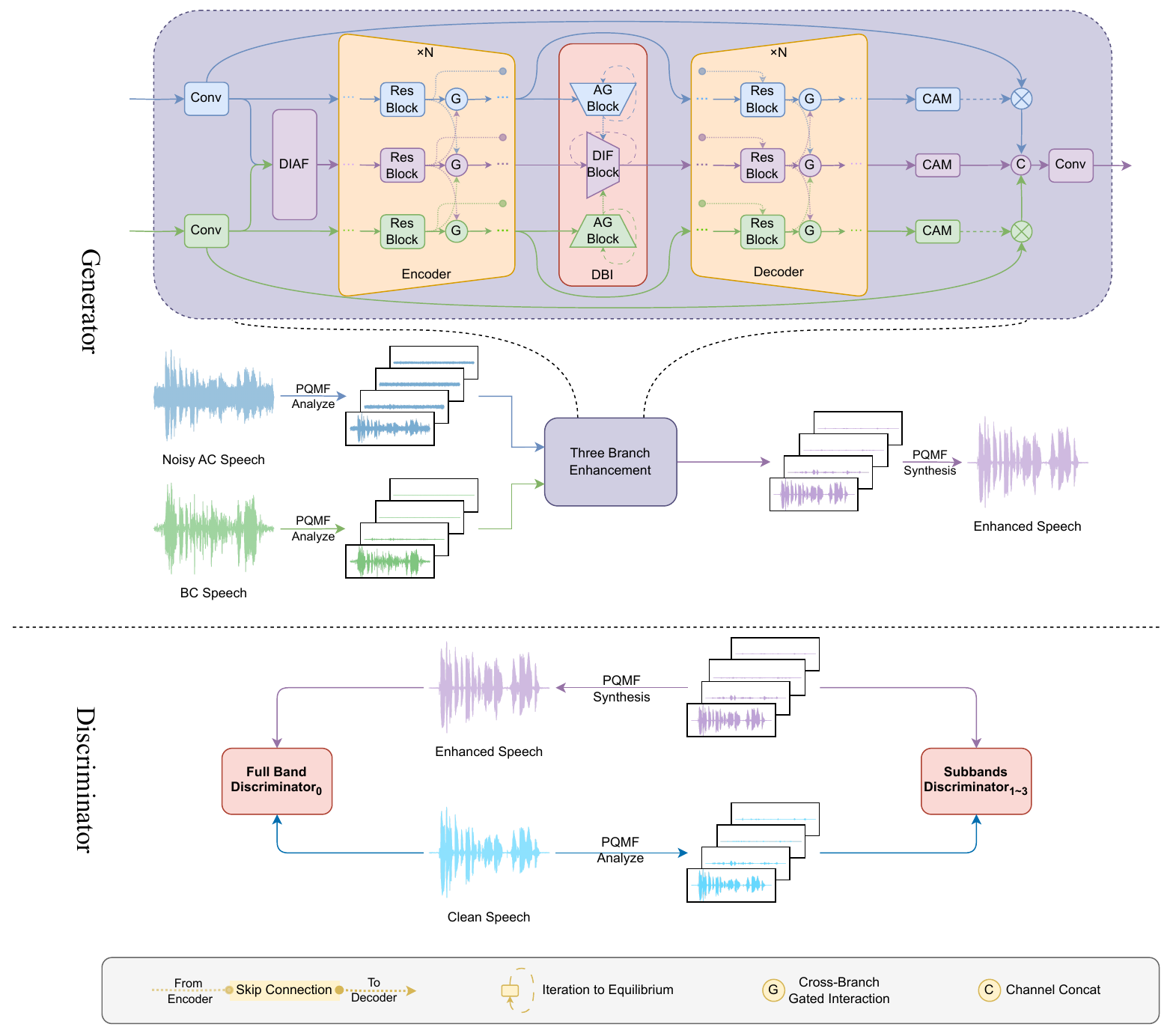}
  \caption{Overall architecture of DBMIF.}
  \label{fig:overview}
\end{figure}

As illustrated in Fig.~\ref{fig:overview}, DBMIF comprises two primary components: a Generator and a multi-scale Discriminator.

The Generator performs the core enhancement task. It initially decomposes the input time-domain speech into four subbands using a PQMF analysis bank, following the configuration recommended in \cite{hauret2023eben}. These subband signals are processed to yield enhanced estimates, which are subsequently reconstructed into a full-band waveform via PQMF synthesis.

Complementing the Generator, the Discriminator adopts a multi-resolution architecture. It scrutinizes not only the reconstructed full-band waveform for global consistency but also the individual subband signals to enforce fine-grained spectral fidelity.

The following sections detail the specific architectures of the Generator and Discriminator, along with the loss functions and training strategy utilized to optimize the framework.

\subsection{Generator}
\label{sec:generator}

\subsubsection{Three-Branch Multi-Scale Interactive Encoder-Decoder}
We design a three-branch architecture consisting of separate AC and BC paths alongside a central fusion branch.
This backbone operates across multiple hierarchical scales to capture diverse spectral contexts. Specifically, we employ progressive channel widths of \{32, 64, 128, 256\} at successive layers, ensuring that the network captures both fine-grained details and high-level semantic abstractions.
The two unimodal branches operate to preserve the discriminative characteristics of each modality, while the fusion stream facilitates continuous information exchange, providing the network with balanced and complementary guidance.

Central to this architecture is the Cross-Branch Gated Interaction (CBGI) module, which enables bidirectional modulation between the fusion branch and the unimodal branches, as illustrated in Fig.~\ref{fig:gate}.
At the $s$-th layer, let $\mathbf{x}_{a}^{(s)}, \mathbf{x}_{b}^{(s)}, \mathbf{x}_{f}^{(s)}$ denote the feature representations of the AC, BC, and fusion branches, respectively.
The CBGI mechanism operates in two distinct stages:

\begin{figure}[ht]
    \centering
    \includegraphics[width=0.65\linewidth]{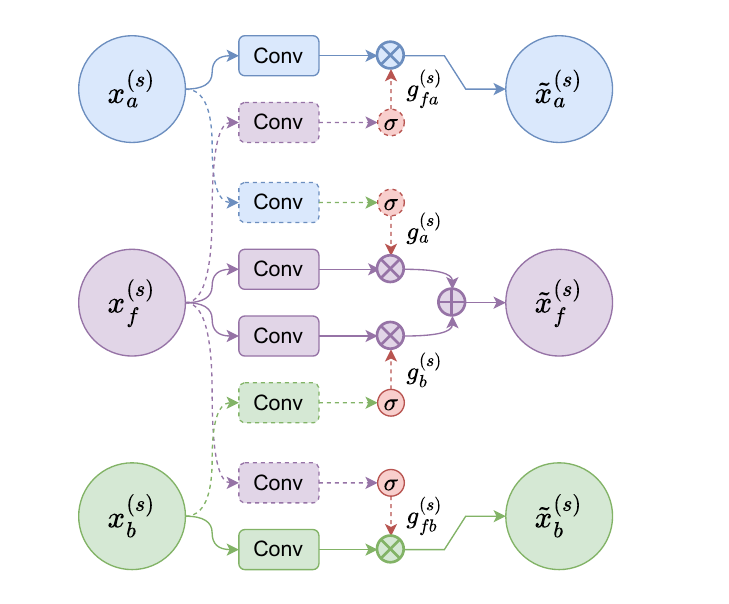}
    \caption{Diagram of the CBGI module. Solid arrows denote the feature flow, while dashed arrows indicate the gating signals generated to modulate the corresponding streams.}
    \label{fig:gate}
\end{figure}

\noindent\textbf{1. Fusion-to-Unimodal Modulation.}
First, the fusion branch generates gating signals to recalibrate the unimodal features. This allows the global context from the fusion stream to suppress irrelevant components in the unimodal branches:
\begin{equation}
\begin{aligned}
g_{fa}^{(s)} &= \sigma\!\bigl(\mathcal{F}_{\mathrm{conv}}(\mathbf{x}_{f}^{(s)})\bigr), \quad & \tilde{\mathbf{x}}_{a}^{(s)} &= g_{fa}^{(s)} \odot \mathcal{F}_{\mathrm{conv}}(\mathbf{x}_{a}^{(s)}), \\
g_{fb}^{(s)} &= \sigma\!\bigl(\mathcal{F}_{\mathrm{conv}}(\mathbf{x}_{f}^{(s)})\bigr), \quad & \tilde{\mathbf{x}}_{b}^{(s)} &= g_{fb}^{(s)} \odot \mathcal{F}_{\mathrm{conv}}(\mathbf{x}_{b}^{(s)}),
\end{aligned}
\end{equation}
where $\mathcal{F}_{\mathrm{conv}}$ denotes a distinct $1\times 1$ convolution layer, $\sigma(\cdot)$ is the sigmoid function, and $\odot$ represents element-wise multiplication. $\tilde{\mathbf{x}}_{a}^{(s)}$ and $\tilde{\mathbf{x}}_{b}^{(s)}$ are the refined unimodal features.

\noindent\textbf{2. Unimodal-to-Fusion Feedback.}
Conversely, to enrich the fusion stream with reliable modality-specific details, each unimodal branch produces feedback gates based on its original features. These gates regulate the integration of the fusion representation:
\begin{equation}
\begin{aligned}
g_{a}^{(s)} &= \sigma\!\bigl(\mathcal{F}_{\mathrm{conv}}(\mathbf{x}_{a}^{(s)})\bigr), \quad & \hat{\mathbf{x}}_{af}^{(s)} &= g_{a}^{(s)} \odot \mathcal{F}_{\mathrm{conv}}(\mathbf{x}_{f}^{(s)}), \\
g_{b}^{(s)} &= \sigma\!\bigl(\mathcal{F}_{\mathrm{conv}}(\mathbf{x}_{b}^{(s)})\bigr), \quad & \hat{\mathbf{x}}_{bf}^{(s)} &= g_{b}^{(s)} \odot \mathcal{F}_{\mathrm{conv}}(\mathbf{x}_{f}^{(s)}).
\end{aligned}
\end{equation}
The updated fusion representation $\tilde{\mathbf{x}}_{f}^{(s)}$ is then obtained by aggregating these gated contributions:
\begin{equation}
\tilde{\mathbf{x}}_{f}^{(s)} = \hat{\mathbf{x}}_{af}^{(s)} + \hat{\mathbf{x}}_{bf}^{(s)}.
\end{equation}

Unlike conventional unidirectional fusion, this bidirectional coupling establishes a closed-loop information flow.
By allowing the fusion features to gate the unimodal branches and vice versa, the mechanism ensures that the network dynamically balances the contributions of both modalities.
This prevents the dominance of a single modality and facilitates the selective integration of high-fidelity components from both streams.

\subsubsection{Early Deep Iterative Attention Fusion}
\label{sec:diaf}

To provide cleaner and more stable inputs for subsequent processing, we employ the Deep Iterative Attention Fusion (DIAF) module at the front end.
This module initializes a coarse fusion and progressively refines it via an attention mechanism.
At each step, a Channel Attention Module (CAM) produces a channel-wise gate that adaptively balances the contributions of each modality, attenuating less reliable modality cues while emphasizing more informative ones.

\begin{figure}[ht]
  \centering
  \includegraphics[width=\linewidth]{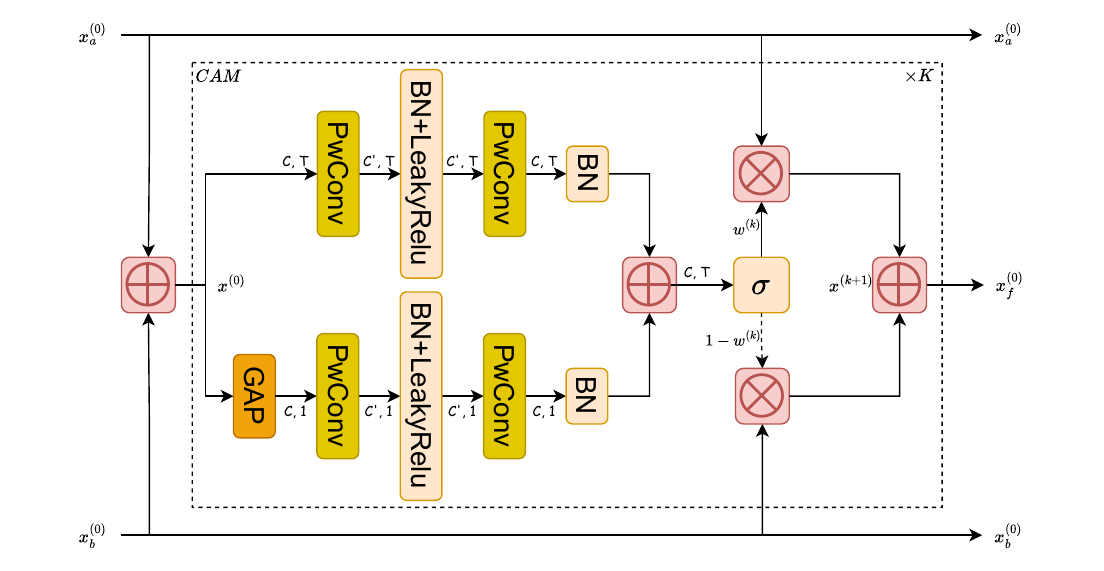}
  \caption{Diagram of the early DIAF module. DIAF progressively refines AC and BC features through iterative channel attention to emphasize reliable modality cues.}
  \label{fig:diaf_overview}
\end{figure}

Let $\mathbf{x}_{a}$ and $\mathbf{x}_{b}$ denote the input feature maps corresponding to the AC and BC speech, respectively.
As illustrated in Fig.~\ref{fig:diaf_overview}, DIAF operates in an iterative manner.
First, to establish a neutral starting point without introducing extra parameters, we initialize the coarse fusion $\mathbf{x}^{(0)}$ as the element-wise sum of these inputs:
\begin{equation}
\mathbf{x}^{(0)} = \mathbf{x}_{a}^{(0)} + \mathbf{x}_{b}^{(0)}.
\end{equation}

Subsequently, for each iteration $k=1,\dots,K$, the CAM generates an attention mask conditioned on the previous fused representation:
\begin{equation}
\mathbf{w}^{(k)} = \operatorname{CAM}\!\big(\mathbf{x}^{(k-1)}\big).
\end{equation}

Here, $\mathbf{w}^{(k)} \in \mathbb{R}^{C}$ serves as a channel-wise reliability gate, with values constrained to $[0,1]$ via a sigmoid nonlinearity following global average pooling.
Mathematically, each element of $\mathbf{w}^{(k)}$ represents the confidence in the AC modality for the corresponding channel, while $1 - \mathbf{w}^{(k)}$ weights the BC modality.
This mask is then applied to refine the fusion through a soft selection mechanism:
\begin{equation}
\mathbf{x}^{(k)} = \mathbf{w}^{(k)} \odot \mathbf{x}_{a}^{(0)} + \big(1-\mathbf{w}^{(k)}\big)\odot \mathbf{x}_{b}^{(0)}.
\end{equation}

After $K$ rounds of refinement, the final quality-aware fusion embedding is obtained as $\mathbf{x}_{f}^{(0)} = \mathbf{x}^{(K)}$.
This iterative process allows the network to re-calibrate modality contributions dynamically as the noise profile changes.
In our implementation, we employ $K=3$, which we found to empirically achieve a robust balance between fusion quality and computational cost.

\subsubsection{Deep Balanced Interaction}
\label{sec:dbi}

The encoder-decoder bottleneck aggregates compact, information-dense features and is thus critical for downstream processing.
To enhance its effectiveness, we employ the Deep Balanced Interaction (DBI) module.
Drawing inspiration from the Deep Equilibrium Multimodal Fusion framework~\cite{ni2023deep}, DBI refines modality-specific features through iterative fixed-point updates.
This equilibrium mechanism stabilizes representations and improves generalization without increasing the network depth or parameter count.
By reusing parameters across iterations, DBI recursively denoises and re-calibrates AC and BC speech features until they reach a balanced state, yielding a consistent representation.

\begin{figure}[ht]
\centering
\begin{subfigure}{0.42\columnwidth}
    \centering
    \includegraphics[width=\linewidth]{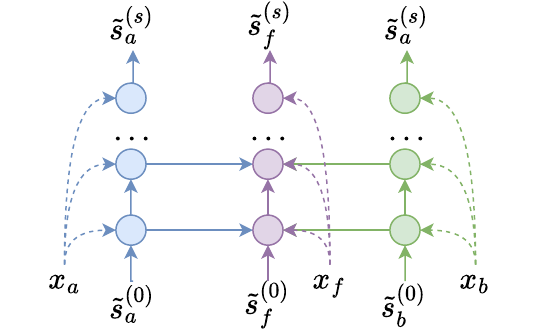}
    \caption{Overall DBI structure}
    \label{fig:dbi_overview}
\end{subfigure}
\hfill
\begin{subfigure}{0.54\columnwidth}
    \centering
    \includegraphics[width=\linewidth]{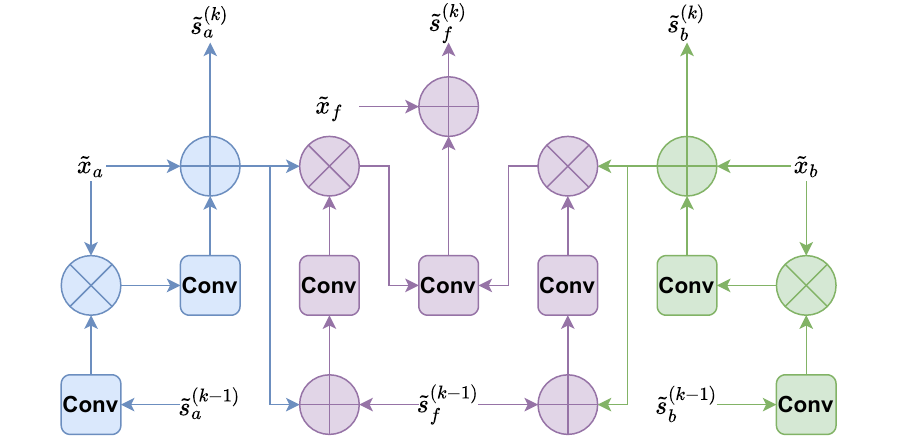}
    \caption{One DBI iteration}
    \label{fig:dbi_iteration}
\end{subfigure}
\caption{
Diagram of DBI module. 
DBI maintains recurrent modality states and iteratively updates them through intra-modal refinement and fusion feedback.
}
\label{fig:dbi}
\end{figure}

As shown in Fig.~\ref{fig:dbi_overview}, the DBI module operates on the bottleneck features from the encoder, denoted as $\mathbf{x}_{a}, \mathbf{x}_{b}, \mathbf{x}_{f}$. These features serve as the time-invariant input injection.
Simultaneously, the module maintains three recurrent hidden states, denoted as $\mathbf{z}_{a}^{(k)}, \mathbf{z}_{b}^{(k)}, \mathbf{z}_{f}^{(k)}$ at iteration $k$.
Fig.~\ref{fig:dbi_iteration} illustrates one update step of the mapping function $\mathcal{F}$, which proceeds in two stages:
\emph{1) Intra-modal refinement:} The states $\mathbf{z}_{a}$ and $\mathbf{z}_{b}$ are updated under the guidance of their corresponding inputs $\mathbf{x}_{a}$ and $\mathbf{x}_{b}$ to suppress noise while preserving modality-specific cues.
\emph{2) Fusion feedback:} The refined unimodal states, along with the previous fusion state $\mathbf{z}_{f}^{(k-1)}$, are integrated under the guidance of $\mathbf{x}_{f}$ to produce the new fusion state $\mathbf{z}_{f}^{(k)}$.
Formally, the joint update rule is defined as:
\begin{equation}
\bigl(\mathbf{z}_{a}^{(k)},\mathbf{z}_{b}^{(k)},\mathbf{z}_{f}^{(k)}\bigr)
= \mathcal{F}\Bigl(\mathbf{z}_{a}^{(k-1)},\mathbf{z}_{b}^{(k-1)},\mathbf{z}_{f}^{(k-1)} \ \Big|\ \mathbf{x}_{a},\mathbf{x}_{b},\mathbf{x}_{f}\Bigr).
\end{equation}

For compact notation, we concatenate the states into a single vector $\mathbf{z}^{(k)}=\bigl[\mathbf{z}_{a}^{(k)},\mathbf{z}_{b}^{(k)},\mathbf{z}_{f}^{(k)}\bigr]$.
Our goal is to apply $\mathcal{F}$ repeatedly such that the sequence converges to a fixed point $\mathbf{z}^\star$ satisfying:
\begin{equation}
\mathbf{z}^\star = \mathcal{F}(\mathbf{z}^\star).
\end{equation}
To improve training stability, we merge the batch and temporal dimensions, treating each frame as an independent sample during the equilibrium process. This allows DBI to adapt to instantaneous changes in modality dominance.

To ensure computational efficiency, we limit the maximum number of iterations to $K_{\max}=50$.
We monitor the convergence by calculating the relative change between consecutive iterates:
\begin{equation}
\delta_k=\bigl\|\mathbf{z}^{(k)}-\mathbf{z}^{(k-1)}\bigr\|_F,\qquad k=2,\dots,K_{\max}.
\end{equation}
Instead of simply taking the last iteration, we select the state $z^\star$ with the smallest change, indicating the most stable equilibrium point, as the final output:
\begin{equation}
k^\star=\arg\min_{2\le k\le K_{\max}}\delta_k,\qquad
\mathbf{z}^\star=\mathbf{z}^{(k^\star)}.
\end{equation}

Since the parameters of $\mathcal{F}$ are shared across all iterations, DBI achieves deep feature refinement and dynamic modality balancing with high parameter efficiency.

\subsection{Discriminator Architecture}
\label{sec:disc_design}

To ensure high-fidelity waveform generation, we employ a multi-scale discrimination strategy inspired by MelGAN~\cite{kumar2019melgan}.
However, unlike conventional multi-scale discriminators that operate solely on waveforms, our design explicitly leverages the subband decomposition to capture fine-grained spectral details.
Specifically, the proposed discriminator consists of two distinct components:

\begin{enumerate}
    \item \textbf{Full-band Waveform Discriminator ($D_{\text{wav}}$):} This component operates on the raw full-band audio at the original sampling rate. It focuses on capturing global structural coherence and low-frequency consistency. As detailed in Table~\ref{tab:dwav_arch}, it is composed of a stack of strided convolutional layers with large receptive fields.
    \item \textbf{Multi-Scale Subband Discriminator ($D_{\text{sub}}$):} We utilize the same PQMF analysis bank employed by the generator to decompose the real or generated speech into 4 subbands. These subbands are concatenated channel-wise and fed into three identical discriminator sub-networks as detailed in Table~\ref{tab:dpqmf_arch}. These sub-networks differ only by their dilation factors $d \in \{1, 2, 3\}$. This dilated convolution design enables the model to capture diverse temporal structures with varying receptive fields within the subbands.
\end{enumerate}

\begin{table}[ht]
\centering
\small
\setlength{\tabcolsep}{4.5pt}
\caption{Architecture of the full-band discriminator $D_{\text{wav}}$. It outputs a time-indexed logit map for global consistency assessment.}
\label{tab:dwav_arch}
\begin{tabular}{c c c c c c c}
\toprule
Layer & $C_{\text{in}}$ & $C_{\text{out}}$ & Kernel & Stride & Groups & Act. \\
\midrule
1 & 1    & 16   & 15 & 1 & 1 & LReLU \\
2 & 16   & 64   & 41 & 4 & 4 & LReLU \\
3 & 64   & 256  & 41 & 4 & 4 & LReLU \\
4 & 256  & 1024 & 41 & 4 & 4 & LReLU \\
5 & 1024 & 1024 & 41 & 4 & 4 & LReLU \\
6 & 1024 & 1024 & 5  & 1 & 1 & LReLU \\
7 & 1024 & 1    & 3  & 1 & 1 & None \\
\bottomrule
\end{tabular}
\end{table}

\begin{table}[ht]
\centering
\small
\setlength{\tabcolsep}{4.5pt}
\caption{Architecture of the subband discriminator $D_{\text{sub}}$. Three copies are instantiated with dilation factors $d \in \{1, 2, 3\}$ to capture multi-scale periodicities in PQMF subbands.}
\label{tab:dpqmf_arch}
\begin{tabular}{c c c c c c c c}
\toprule
Layer & $C_{\text{in}}$ & $C_{\text{out}}$ & Kernel & Stride & Groups & Dilation & Act. \\
\midrule
1 & 4    & 36    & 3 & 1 & 4 & $d$ & LReLU \\
2 & 36   & 72    & 7 & 2 & 4 & $d$ & LReLU \\
3 & 72   & 144   & 7 & 2 & 4 & $d$ & LReLU \\
4 & 144  & 288   & 7 & 2 & 4 & $d$ & LReLU \\
5 & 288  & 576   & 7 & 2 & 4 & $d$ & LReLU \\
6 & 576  & 1152  & 7 & 2 & 4 & $d$ & LReLU \\
7 & 1152 & 1152  & 5 & 1 & 4 & $d$ & LReLU \\
8 & 1152 & 1     & 3 & 1 & 1 & $d$ & None \\
\bottomrule
\end{tabular}
\end{table}

It should be noted that, all convolutional layers in both discriminators apply weight normalization and use LeakyReLU activation ($\alpha=0.1$), except for the final layer which outputs a linear score map.

\subsection{Training Objectives}
\label{sec:disc_train}

The optimization of the proposed framework follows an adversarial learning paradigm, involving a minimax game between a multi-scale discriminator and a generator. The training process is governed by two distinct yet complementary objectives: the discriminator is trained to construct a robust classification boundary between natural and synthesized waveforms, while the generator aims to synthesize perceptually realistic speech that successfully deceives the discriminator. In the following subsections, we detail the optimization targets for the discriminator and the generator, respectively.

\subsubsection{Discriminator Loss}
\label{sec:loss_d}
We employ a hinge loss formulation for the adversarial training of the discriminators, as it provides a robust gradient signal compared to standard cross-entropy objectives. The total discriminator loss $L_D$ aggregates the contributions from a full-band discriminator and three subband discriminators (indexed by $k=0, \dots, 3$). This multi-resolution evaluation scheme enables the model to simultaneously supervise global spectral envelopes and fine-grained local textures. Let $D_k(\mathbf{x})$ denote the output score map of the $k$-th discriminator for an input $\mathbf{x}$. The discriminators are optimized to distinguish between the ground-truth speech $\mathbf{y}$ and the synthesized speech $\hat{\mathbf{y}}$ by minimizing:
\begin{equation}
L_D = \sum_{k=0}^{3} \left[
\mathbb{E}_{\mathbf{y}} \bigl[\max(0, 1 - D_k(\mathbf{y}))\bigr] +
\mathbb{E}_{\hat{\mathbf{y}}} \bigl[\max(0, 1 + D_k(\hat{\mathbf{y}}))\bigr]
\right].
\label{eq:disc_loss}
\end{equation}

\subsubsection{Generator Loss}
\label{sec:loss_g}
The generator aims to minimize a composite objective that balances perceptual realism with structural fidelity. To drive the synthesis of realistic waveform structures, we first define the adversarial component $L_G^{\mathrm{adv}}$ using the hinge loss. By encouraging the generated samples to achieve scores exceeding the unit margin, the generator is penalized for any artifacts detected by the discriminators:
\begin{equation}
L_G^{\mathrm{adv}} = \sum_{k=0}^{3} \mathbb{E}_{\hat{\mathbf{y}}} \bigl[ \max(0, 1 - D_k(\hat{\mathbf{y}})) \bigr].
\label{eq:gen_adv_loss}
\end{equation}

To complement the adversarial objective and mitigate the over-smoothing often observed in GAN-based synthesis, we incorporate a feature matching loss $L_G^{\mathrm{fm}}$. This term stabilizes training by minimizing the $L_1$ distance between the intermediate feature maps of the discriminators for real and generated inputs, effectively acting as a learned perceptual metric:
\begin{equation}
L_G^{\mathrm{fm}} = \sum_{k=0}^{3} \sum_{i=1}^{N_k-1} \frac{1}{M_i} \left\| D^{(i)}_k(\mathbf{y}) - D^{(i)}_k(\hat{\mathbf{y}}) \right\|_1,
\label{eq:gen_fm_loss}
\end{equation}
where $D^{(i)}_k$ denotes the feature map extracted from the $i$-th layer of the $k$-th discriminator, and $M_i$ serves as a normalization factor representing the number of elements in that layer.

Consequently, the total training objective for the generator is formulated as a weighted combination of these two terms:
\begin{equation}
L_G = L_G^{\mathrm{adv}} + \lambda L_G^{\mathrm{fm}},
\label{eq:gen_total_loss}
\end{equation}
where $\lambda$ is a hyperparameter that controls the relative importance of the feature matching supervision.

\section{Experimental Setup}

\subsection{Datasets}
We use the Air- and Bone-Conducted Synchronized Speech corpus~\cite{wang2022end} as our dataset. The corpus contains time-aligned AC and BC speech recordings from 100 speakers which comprises 47,182 utterances and 42 hours in total, with durations ranging from 1 to 5 seconds. All audio recordings are sampled at 16~kHz.

During training, we employ an online slicing and noise-mixing pipeline where each utterance is segmented into 1-second windows. We inject noise exclusively into the AC speech channel using samples sourced from the DNS Challenge~\cite{reddy2021dnsmos} and UTD Environmental Noise~\cite{saki2016smartphone} datasets. These noise sources cover diverse categories such as machinery, in-car environments, and babble. The SNR is uniformly sampled from the range $[-15, 5]$~dB. Following noise injection, we rescale the mixture to match the Root Mean Square energy of the original clean speech. Finally, we apply a 50~ms linear fade-in and fade-out to the segment boundaries to mitigate discontinuities caused by slicing.

For evaluation, we utilize three unseen noise corpora to ensure diverse acoustic conditions. These include QUT-NOISE~\cite{dean2010qut} for realistic scenarios such as cafés and cars, NOISEX-92~\cite{varga1993assessment} for standardized industrial and military benchmarks, and Nonspeech~\cite{hu2010tandem} for everyday ambient noises. Distinct from the training phase, we process the test utterances in their full length without segmentation. Each test utterance is mixed with these noise sources under fixed conditions where the SNR is selected from the set ({-15, -10, -5, 0, 5})~dB.

\subsection{Comparison Baselines}
To comprehensively evaluate the effectiveness of DBMIF, we benchmark it against representative baselines as summarized in Table~\ref{tab:modelspecs}. These baselines primarily consist of multimodal approaches, including time-domain models such as FCN~\cite{yu2020time} and MMINet~\cite{wang2022multi}, as well as time–frequency domain models such as DenGCAN~\cite{kuang2024lightweight} and Dc--Crn~\cite{wang2022fusing}. Furthermore, to verify the benefits of multimodal fusion, we also incorporate unimodal baselines, specifically GaGNet~\cite{li2022glance} for AC speech enhancement and EBEN~\cite{hauret2023eben} for BC speech enhancement.

\begin{table}[ht]
\centering
\caption{Details of compared models.}
\label{tab:modelspecs}
\begin{tabular}{lccc}
\toprule
Model & Modality & Domain & GAN-based \\
\midrule 
GaGNet~\cite{li2022glance} & AC & Time--Frequency & No \\
EBEN~\cite{hauret2023eben} & BC & Time & Yes \\
FCN~\cite{yu2020time}        & AC\&BC & Time              & No \\
MMINet~\cite{wang2022multi}  & AC\&BC & Time              & No \\
DenGCAN~\cite{kuang2024lightweight} & AC\&BC & Time--Frequency & No \\
Dc--Crn~\cite{wang2022fusing} & AC\&BC & Time--Frequency  & No \\
\textbf{DBMIF (ours)}        & AC\&BC & Time & \textbf{Yes} \\
\bottomrule
\end{tabular}
\end{table}

\subsection{Configuration and Implementation Details}
The model is trained using synchronized 1 second AC and BC speech pairs with a batch size of 32 for a total of 100 epochs. Optimization is performed using the Adam optimizer with a learning rate of \(3\times10^{-4}\) and momentum parameters \((\beta_{1}, \beta_{2}) = (0.5, 0.9)\). A cosine annealing schedule is employed to gradually decay the learning rate throughout training. During adversarial learning, the generator and discriminator are updated alternately, and the feature-matching loss is weighted by \(\lambda = 1000\).

\subsection{Evaluation Metrics}
Performance is evaluated using five widely adopted objective metrics, including scale-invariant signal-to-distortion ratio (SI-SDR), perceptual evaluation of speech quality (PESQ)~\cite{rix2001perceptual}, short-time objective intelligibility (STOI)~\cite{taal2011stoi}, extended STOI (ESTOI)~\cite{jensen2016algorithm}, and DNSMOS~\cite{reddy2021dnsmos}. Unlike the first four intrusive metrics that require a clean reference, DNSMOS is a non-intrusive deep neural model designed to estimate the perceptual quality of speech directly from noisy or enhanced signals. It has been shown to correlate well with human mean opinion scores, making it suitable for large-scale evaluation without subjective listening tests. 

To further assess the practical benefits of the enhanced speech, we perform downstream ASR experiments. CER is reported using two representative end-to-end ASR frameworks, Whisper~\cite{radford2023robust} and WeNet~\cite{yao2021wenet}. Whisper offers strong multilingual and noise-robust recognition capabilities, while WeNet is a lightweight, open-source streaming ASR framework optimized for efficiency in specific language settings.

\section{Experimental Results}

We present a comprehensive evaluation of the proposed DBMIF framework to validate its effectiveness and robustness. We first benchmark DBMIF against representative unimodal and multimodal baselines, analyzing performance through objective metrics, downstream ASR accuracy, and subjective listening tests. To assess generalization capabilities, we further examine performance consistency across varying SNR conditions and diverse noise categories, supported by detailed spectral analysis. Finally, we conduct systematic ablation studies to verify the contributions of key architectural components and visualize the adaptive behaviors of the modality fusion mechanism.

\subsection{Comparison with Baselines}
Table~\ref{tab:overall} presents a quantitative performance summary across all models. The unprocessed AC speech suffers from severe noise degradation and is characterized by poor signal integrity with an SI-SDR of -4.96~dB and a high Whisper ASR error rate of 47.79\%. Unimodal baselines exhibit distinct limitations. Specifically, the AC-based GaGNet achieves only marginal improvements in PESQ, while the BC-based EBEN is constrained by limited bandwidth and yields extremely low SI-SDR scores.

\subsubsection{Overall Quantitative Comparison}
\begin{table}[ht]
\centering
\caption{Comparison of objective metrics between raw signals and enhanced speech.}
\label{tab:overall}
\scriptsize 
\begin{tabular}{lccccccc}
\toprule
Model & SI--SDR & PESQ & STOI & ESTOI & DNSMOS & Whisper & WeNet \\
\midrule
\multicolumn{8}{l}{\emph{Origin}} \\
\quad AC speech & -4.96 & 1.13 & 0.68 & 0.50 & 2.39 & 47.79 & 34.62 \\
\quad BC speech & -7.53 & 1.43 & 0.69 & 0.58 & 2.48 & 29.30 & 25.44 \\
\midrule
\multicolumn{8}{l}{\emph{AC-only}} \\
\quad GaGNet~\cite{li2022glance}    & 7.84 & 1.55 & 0.77 & 0.63 & 2.69 & 40.17 & 26.17 \\
\midrule
\multicolumn{8}{l}{\emph{BC-only}} \\
\quad EBEN~\cite{hauret2023eben}    &  0.80 & 1.84 & 0.79 & 0.70 & 3.17 & 43.86 & 34.72 \\
\midrule
\multicolumn{8}{l}{\emph{Joint Model}} \\
\quad FCN~\cite{yu2020time}     &  8.16 & 1.43 & 0.77 & 0.66 & 2.39 & 50.83 & 29.65 \\
\quad MMINet~\cite{wang2022multi}  & 10.45 & 1.77 & 0.83 & 0.73 & 2.69 & 32.58 & 24.65 \\
\quad DenGCAN~\cite{kuang2024lightweight}  & 10.75 & 1.98 & 0.88 & 0.81 & 2.94 & 20.65 & 14.80 \\
\quad Dc--Crn~\cite{wang2022fusing} & 10.18 & 2.62 & 0.86 & 0.79 & 2.66 & 19.46 & 14.50 \\
\quad \textbf{DBMIF} & 12.42 & 2.70 & 0.90 & 0.85 & 3.35 & 15.84 & 12.07 \\
\bottomrule
\end{tabular}
\end{table}

Generally, fusion-based baselines achieve improved signal quality metrics compared to the original noisy inputs which verifies the effectiveness of the multimodal approach. However, distinct performance gaps exist across architectures. Time-domain models such as FCN and MMINet perform poorly on ASR tasks despite gains in other metrics. This discrepancy likely arises because full-band time-domain convolution can introduce structural artifacts or over-smooth high-frequency details which disrupt the acoustic cues required for machine recognition. In contrast, T-F methods including DenGCAN and Dc-CRN achieve more balanced performance across validity and intelligibility but still leave a non-trivial gap relative to clean speech. 
Finally, DBMIF achieves state-of-the-art results across all metrics. This superior performance is likely attributed to the PQMF-based subband strategy and the sophisticated interaction mechanism between the AC and BC speech branches, which together mitigate artifacts common in conventional full-band time-domain modeling.

\subsubsection{Performance Consistency Across SNR Regimes}

\begin{figure*}[ht]
  \centering
  \includegraphics[width=\linewidth]{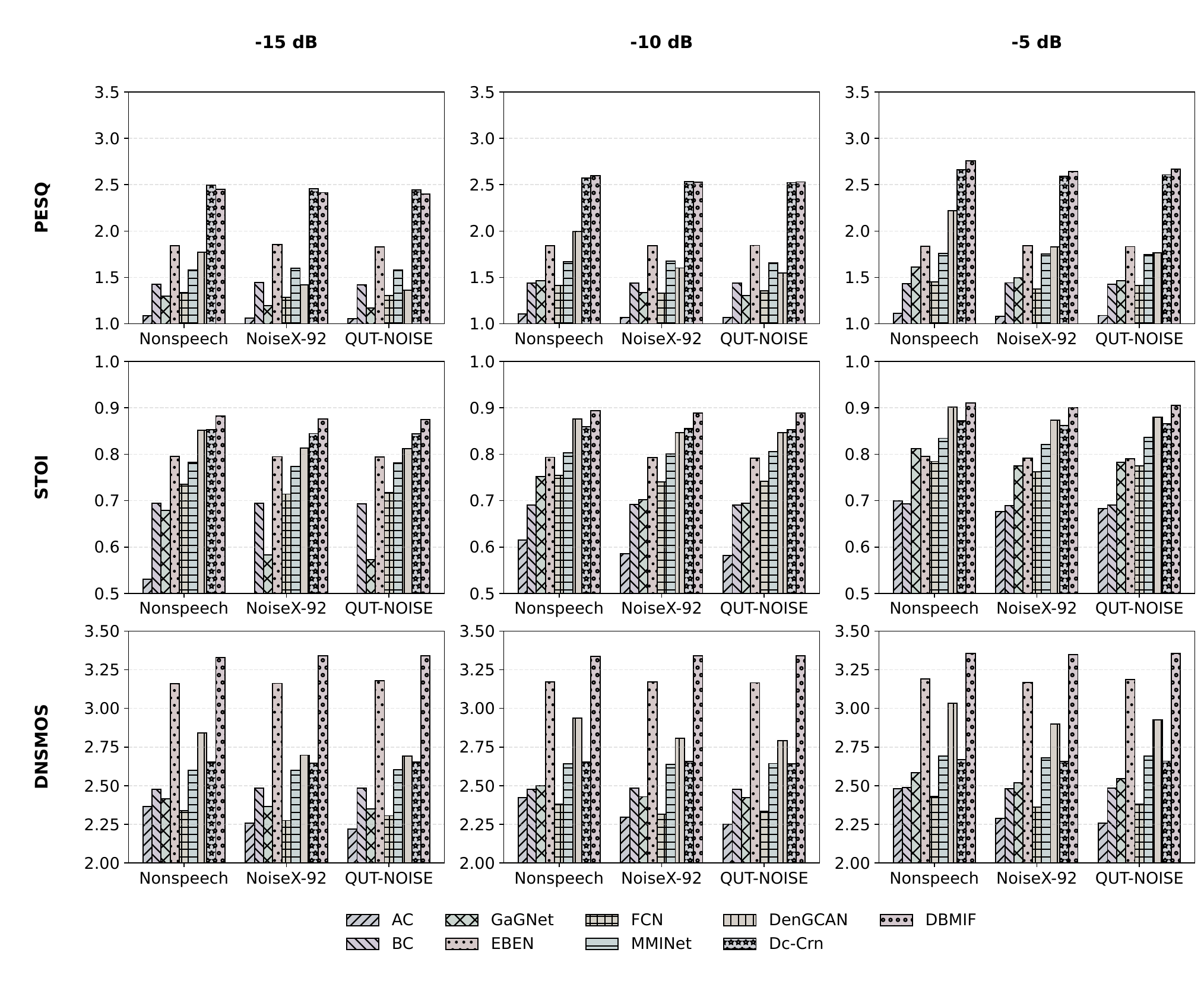} 
  \caption{Low-SNR results among different methods.}
  \label{fig:low_snr}
\end{figure*}

\begin{figure*}[ht]
  \centering
  \includegraphics[width=0.70\linewidth]{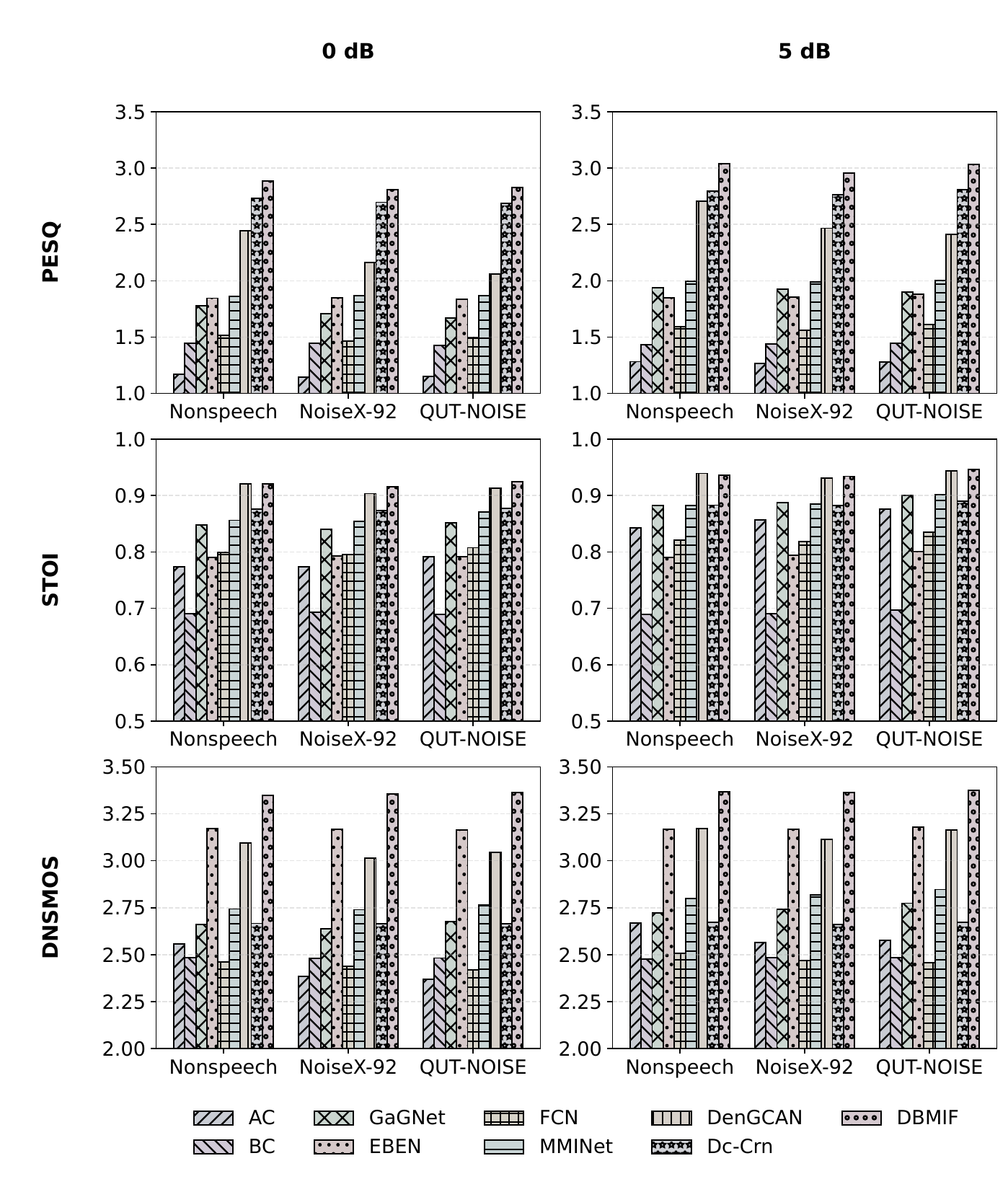} 
  \caption{High-SNR results among different methods.}
  \label{fig:high_snr}
\end{figure*}

We compare methods across various noise corpora under low SNR conditions at -15, -10, and -5 dB. 
As shown in Fig.~\ref{fig:low_snr}, GaGNet degrades severely at low SNRs and fluctuates widely across noise types. 
DenGCAN also struggles, likely due to limited capacity for complex non-stationary noise. In contrast, the remaining multimodal approaches are more robust across datasets, indicating that incorporating BC speech mitigates degradation under unseen noise. 
Moreover, both EBEN and DBMIF use GAN training and obtain higher DNSMOS, which suggests adversarial learning improves perceptual quality and naturalness. 
Overall, DBMIF yields consistent improvements across metrics in low-SNR regimes and typically ranks among the top-performing methods across noise conditions. We also observe a few extreme cases where the advantage becomes smaller on certain perceptual metrics. For instance, at -15 dB in highly non-stationary babble noise, the BC-only baseline EBEN achieves a DNSMOS score comparable to DBMIF. This is attributed to the inherent physical characteristics of BC sensors, which produce noise-free but bandwidth-limited signals. In contrast, DBMIF prioritizes the restoration of high-frequency details. Consequently, DBMIF yields significantly more natural speech, even if it entails retaining minimal residual noise.

Fig.~\ref{fig:high_snr} presents a performance comparison under high-SNR conditions at 0 and 5 dB.
As the SNR increases and the quality of the AC speech improves, methods integrating the AC speech modality demonstrate consistent performance enhancements. Notably, at an SNR of $5$~dB, these frameworks surpass methods relying solely on BC speech in terms of PESQ and STOI. This trend underscores the inherent bandwidth limitations of the BC speech modality in scenarios where clean acoustic features from the AC speech are available.

Among the evaluated baselines, MMINet maintains a steady performance baseline across diverse noise environments, whereas DenGCAN, which appears vulnerable in low-SNR conditions, shows a substantial recovery as the acoustic environment clears. 
Most notably, DBMIF consistently ranks among the top-performing methods across all tested scenarios. 

Overall, DBMIF demonstrates robust stability across the entire SNR range, consistently ranking among the top-performing methods in all evaluated metrics. Meanwhile, it displays strong generalization capabilities across diverse noise conditions.

\subsubsection{Analysis of the Spectrogram}

\begin{figure*}[ht]
  \centering
  \includegraphics[width=\linewidth]{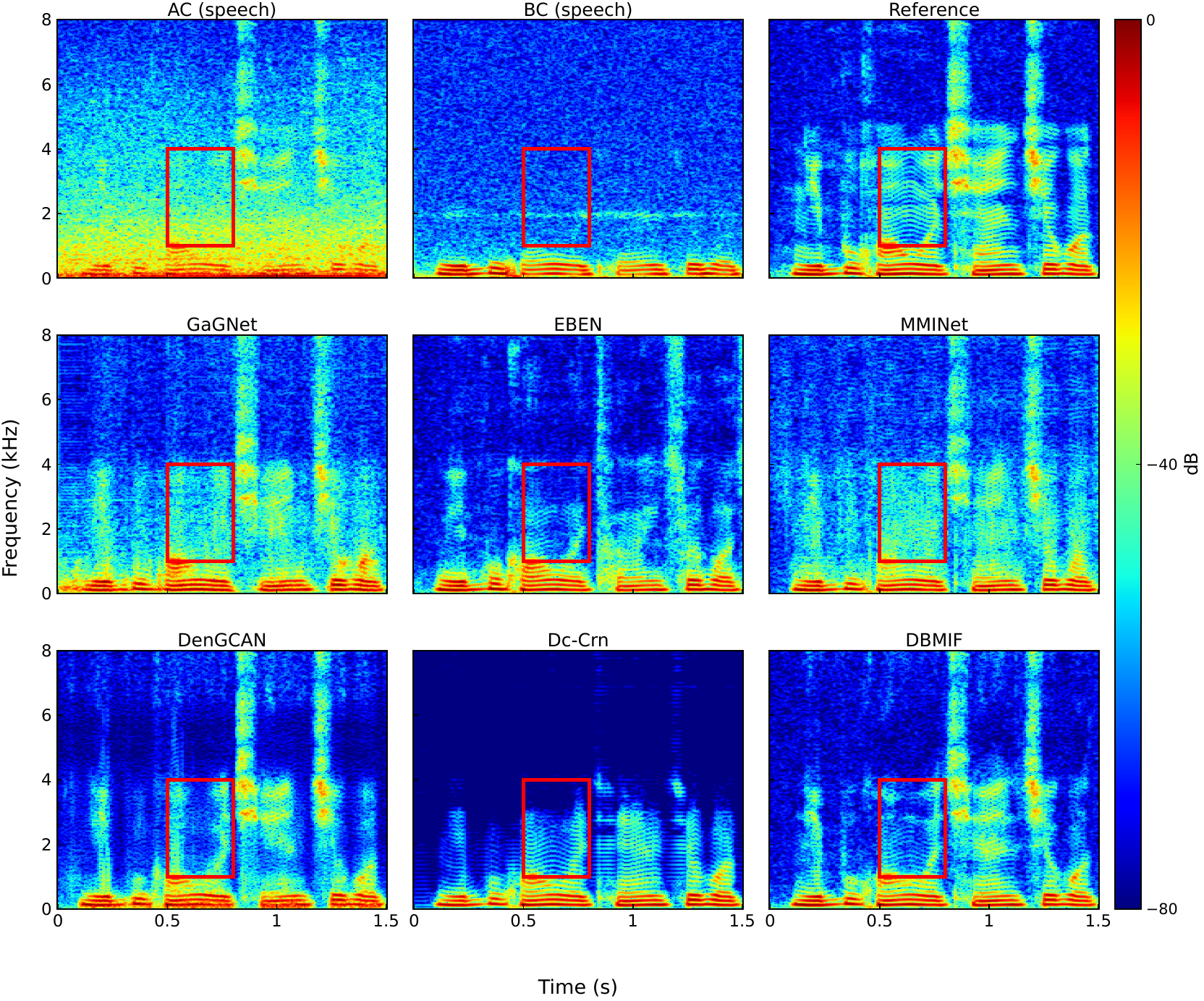}
  \caption{Spectrogram comparison at -10~dB across methods.}
  \label{fig:spec_-10dB}
\end{figure*}

Fig.~\ref{fig:spec_-10dB} presents the spectrogram comparisons in the presence of street noise at an SNR of $-10$~dB. Under such severe noise conditions, the AC speech signal is heavily masked by background noise, whereas the BC speech preserves essential components primarily within the low-frequency region below 1~kHz. While GaGNet achieves partial noise suppression, it introduces noticeable artifacts. Meanwhile, EBEN produces a cleaner output but notably lacks sufficient high-frequency content. Among multi-modal approaches, Dc--Crn preserves low-frequency energy but fails to effectively reconstruct mid-to-high-frequency details, resulting in perceptible distortion. Similarly, although MMINet and DenGCAN aim to restore high-frequency components, MMINet yields overly smoothed spectra, and DenGCAN offers minimal spectral improvement, appearing blurred. In contrast, DBMIF achieves the most significant reconstruction fidelity, with its spectral patterns closely aligning with the clean reference in the highlighted regions. It effectively eliminates the structural artifacts observed in Dc--Crn and recovers fine spectral details missed by other models. These findings demonstrate that DBMIF successfully leverages the low-frequency guidance from BC speech while retaining the high-frequency and fine-grained features of AC speech, thereby ensuring superior naturalness and intelligibility in adverse acoustic scenarios.

\subsubsection{Subjective Evaluation}

\begin{figure}[ht]
\centering
\includegraphics[width=\linewidth]{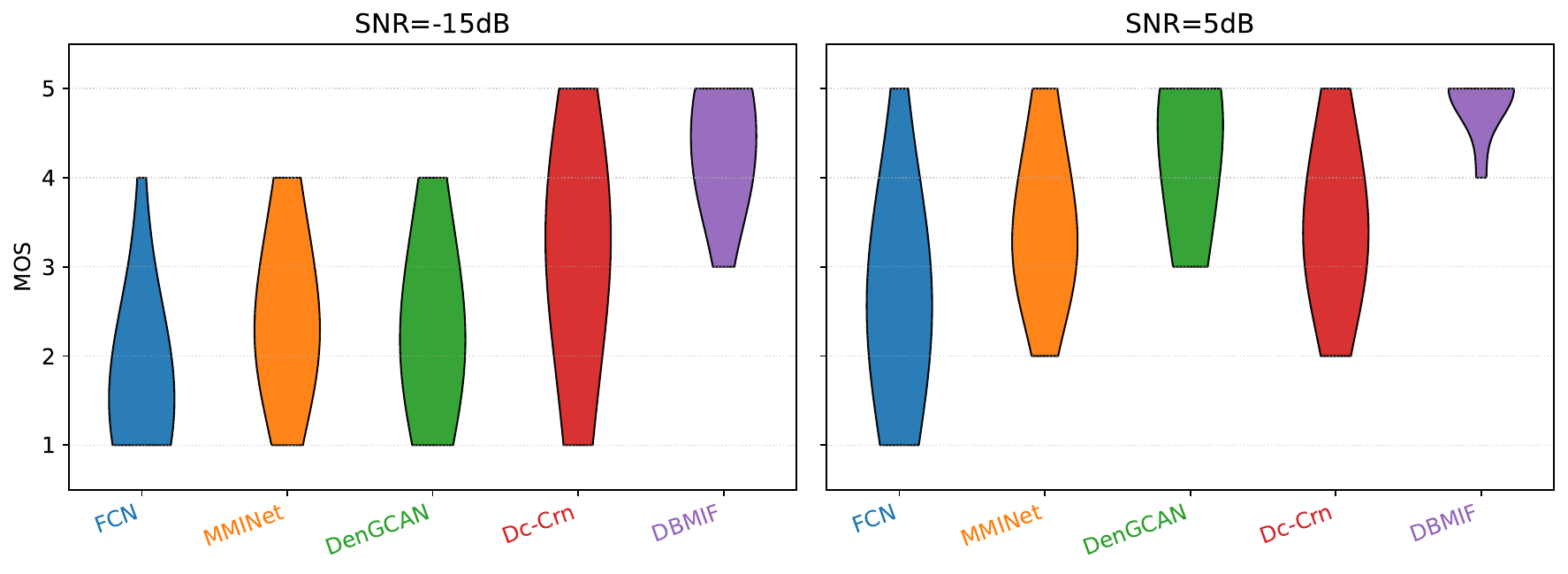}
\caption{Mean opinion scores obtained from the subjective listening test. Wider sections represent higher score density, with higher positions indicating better perceptual quality.}
\label{fig:subjective}
\end{figure}

To strictly evaluate perceptual quality, we conducted subjective listening tests adopting the absolute category rating protocol. Ten native Chinese speakers participated. Clean AC speech and unprocessed noisy signals served as anchors to calibrate the rating scale. Fig.~\ref{fig:subjective} visualizes the distribution of MOS under extreme noise conditions. At the severe SNR of $-15$~dB, comparative methods exhibit wide variances with scores frequently dropping to 1, indicating significant signal degradation. In contrast, DBMIF demonstrates superior robustness, its MOS ratings are densely clustered between 4 and 5, with no samples scored below 3. This suggests that DBMIF effectively utilizes the BC signal to prevent intelligibility collapse even when the AC signal is heavily corrupted. At $5$~dB, DBMIF further distinguishes itself through high consistency. Unlike baseline methods that still retain a tail of lower scores due to residual artifacts, DBMIF maintains a compact distribution near the maximum MOS, confirming its ability to preserve speech naturalness and fine details without introducing audible distortions.

\subsection{Ablation Studies}

\subsubsection{Impact of Architectural Components}

\begin{table*}[ht]
\centering
\begin{threeparttable}
\caption{Ablation study on architectural components across three branches.}
\label{tab:ablation_modules}
\begin{tabular}{ccc cc}
\toprule
\multicolumn{3}{c}{\textbf{Components}} & \multicolumn{2}{c}{\textbf{CER (\%) ↓\tnote{a}}} \\
\cmidrule(lr){1-3}\cmidrule(lr){4-5}
\textbf{CBGI module} & \textbf{DIAF} & \textbf{DBI} & \textbf{Whisper (ASR)} & \textbf{WeNet (ASR)} \\
\midrule
\multicolumn{3}{l}{AC Speech} & 47.79 & 34.62 \\
\multicolumn{3}{l}{BC Speech} & 29.30 & 25.44 \\
\midrule
\multicolumn{3}{l}{Three-Branch Baseline } & 26.90 & 22.00 \\
\checkmark & -- & -- & 19.80 & 15.25 \\
\checkmark & \checkmark & -- & 18.50 & 14.68 \\
\checkmark & \checkmark & \checkmark & 15.84 & 12.77 \\
\bottomrule
\end{tabular}
\begin{tablenotes}
    \small
    \item[a] The down arrow (↓) indicates that lower values represent better performance.
\end{tablenotes}
\end{threeparttable}
\end{table*}

Table~\ref{tab:ablation_modules} presents the cumulative ablation studies evaluating the contribution of each component.
The baseline is defined as a standard three-branch model excluding the CBGI, DIAF, and DBI modules, with each component added incrementally to assess its individual impact.
The proposed architecture achieves substantial gains in downstream ASR performance across both Whisper and WeNet backends.
Notably, the CBGI module yields the most significant improvement, reducing the CER by approximately 6\% for each ASR system.
This highlights its critical role in enabling gated information exchange while preserving modality-specific representations.
In our preliminary experiments, we implemented a standard Gated Multimodal Unit (GMU), but observed suboptimal performance compared to the proposed CBGI.
Additionally, the DIAF and DBI modules provide complementary benefits by refining the fusion process and strengthening inter-branch cooperation, leading to further performance gains.
Overall, these results demonstrate that the full combination of modules achieves the most effective integration of complementary cues from AC and BC speech, resulting in a robust multimodal representation.

\subsubsection{Impact of Branching Strategies}

We evaluate single-, dual-, and three-branch variants to examine how branching configurations affect performance under varying noise conditions.
In the single-branch variant, AC and BC speech features are fused early by DIAF and processed by a shared encoder–decoder, representing a simplified single-stage design without explicit modality separation.
The dual-branch variant processes modalities independently with cross-gating attention mechanisms but lacks a dedicated final fusion stream.
The three-branch configuration represents our full proposed model, where an additional branch is dedicated to integrating complementary cues from both modalities.

\begin{table}[ht]
\centering
\caption{Performance by SNR for branching strategies.}
\label{tab:ablation_snr}
\begin{tabular}{lccc ccc ccc}
\toprule
& \multicolumn{3}{c}{\textbf{SI--SDR}} & \multicolumn{3}{c}{\textbf{STOI}} & \multicolumn{3}{c}{\textbf{DNSMOS}} \\
\cmidrule(lr){2-4}\cmidrule(lr){5-7}\cmidrule(lr){8-10}
Model & $-15$ & $-5$ & $5$ & $-15$ & $-5$ & $5$ & $-15$ & $-5$ & $5$ \\
\midrule
AC Speech        & -14.71 & -4.73 & 5.03 & 0.52 & 0.69 & 0.85 & 2.29 & 2.48 & 2.60 \\
Single-branch   & 0.86 & 6.04 & 10.96 & 0.67 & 0.75 & 0.81 & 2.57 & 2.66 & 2.74 \\
Dual-branch     & 7.15 & 9.98 & 11.96 & 0.82 & 0.86 & 0.91 & 3.19 & 3.21 & 3.23 \\
Three-branch    & 9.27 & 12.86 & 15.89 & 0.88 & 0.91 & 0.94 & 3.34 & 3.35 & 3.36 \\
\bottomrule
\end{tabular}
\end{table}

Results in Table~\ref{tab:ablation_snr} reveal significant performance disparities among the strategies.
The single-branch baseline performs the worst, particularly at low SNR conditions, indicating that simple early fusion struggles to disentangle clean speech from heavy noise due to modality dominance issues.
The dual-branch configuration consistently outperforms the single-branch setup, demonstrating the benefit of maintaining independent representations with mutual guidance.
The three-branch model achieves the best results across all metrics. The dedicated fusion branch effectively strengthens cross-modal exchange and suppresses artifacts, yielding clearer speech.

Interestingly, DNSMOS remains consistently high across SNRs due to metric saturation. The model leverages BC cues to suppress dominant artifacts even at $-15$ dB, ensuring base perceptual quality. While recovering AC speech details at higher SNRs significantly enhances fidelity, it yields diminishing perceptual gains.

\begin{table}[ht]
\centering
\caption{Performance by noise category for branching strategies(Impulsive, Stationary, and Babble)}
\label{tab:ablation_noise}
\begin{tabular}{lccc ccc ccc}
\toprule
& \multicolumn{3}{c}{\textbf{SI--SDR}} & \multicolumn{3}{c}{\textbf{STOI}} & \multicolumn{3}{c}{\textbf{DNSMOS}} \\
\cmidrule(lr){2-4}\cmidrule(lr){5-7}\cmidrule(lr){8-10}
Model & Imp. & Sta. & Bab. & Imp. & Sta. & Bab. & Imp. & Sta. & Bab. \\
\midrule
AC Speech        & -4.87 & -4.71 & -4.82 & 0.60 & 0.66 & 0.60 & 2.16 & 2.23 & 2.33 \\
Single-branch   & 4.62 &  7.39 & 4.59 & 0.74 & 0.76 & 0.75 & 2.67 & 2.62 & 2.68 \\
Dual-branch     & 8.92 &  9.53 & 9.04 & 0.84 & 0.83 & 0.85 & 3.22 & 3.20 & 3.21 \\
Three-branch    & 11.26 & 12.41 & 11.24 & 0.89 & 0.89 & 0.90 & 3.34 & 3.30 & 3.38 \\
\bottomrule
\end{tabular}
\end{table}

Table~\ref{tab:ablation_noise} further compares performance across specific noise categories.
The single-branch model exhibits significant performance drops under impulsive and babble noise, revealing its limitation in handling abrupt transients or complex speech-like interference.
In contrast, the dual- and three-branch configurations show consistent resilience, validating that BC speech provide reliable cues when AC speech is severely degraded.
The three-branch design further amplifies this advantage through its gated fusion mechanism, which selectively suppresses unreliable information flow.
Consequently, as the reliability of AC speech decreases in challenging noise environments, the structural advantage of the three-branch integration becomes increasingly pronounced.

\subsubsection{Adaptive Modality Fusion Behavior in DIAF}

\begin{figure}[ht]
\centering
\includegraphics[width=\linewidth]{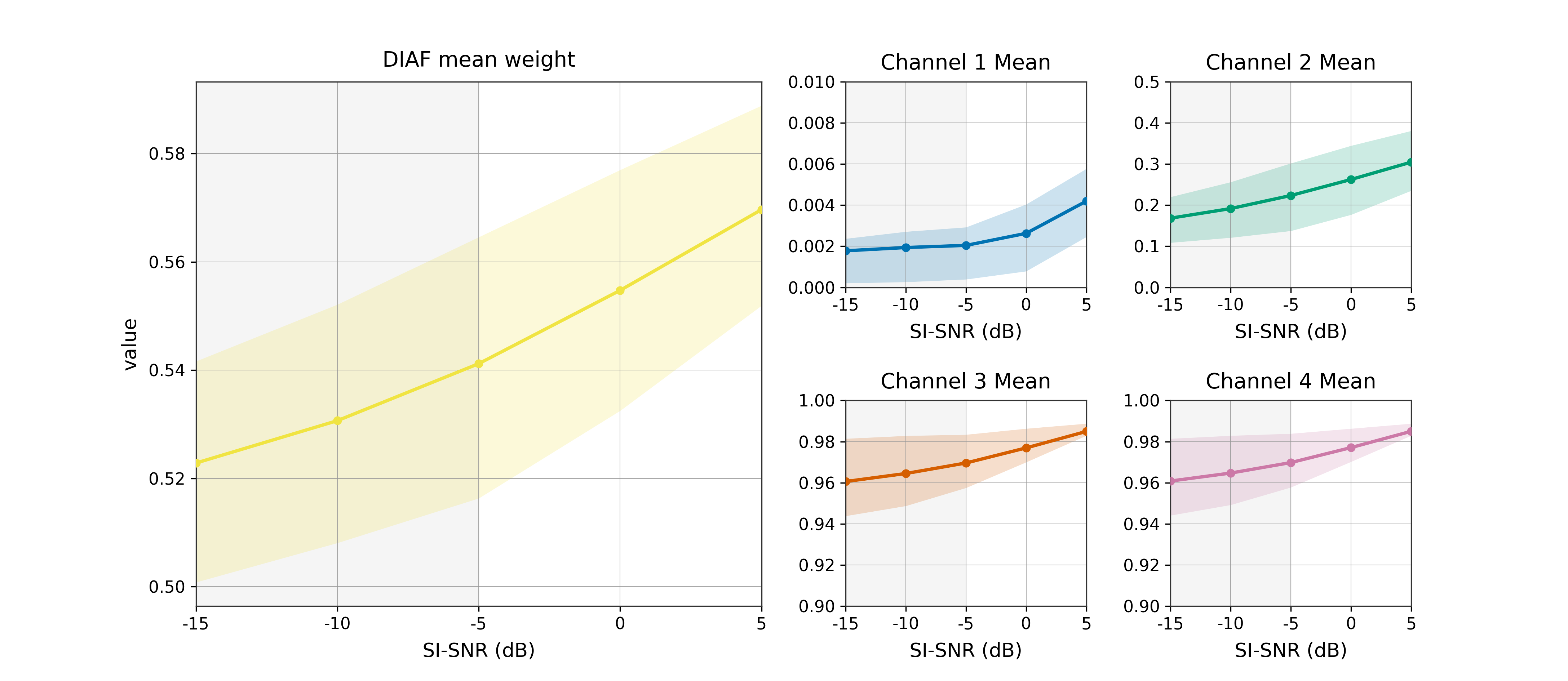}
\caption{DIAF fusion weights $w^{(3)}$ ($w \to 1$ implies AC reliance). Left: Global mean vs. SNR with shaded interquartile range. Right: Frequency-dependent adaptation across sub-bands.}
\label{fig:gamma_snr}
\end{figure}

Fig.~\ref{fig:gamma_snr} visualizes the evolution of the fusion weight $w^{(3)}$, where a higher value indicates greater reliance on the AC modality. As shown in the left panel, the mean weight exhibits a positive correlation with SNR, confirming that DIAF is SNR-aware: it prioritizes the noise-immune BC signal in low-SNR regimes and progressively shifts toward AC as conditions improve to recover spectral fidelity.The right panel further reveals a frequency-dependent adaptation strategy. For low frequencies (Channel 1), the model consistently maintains low weights to anchor pitch information on the robust BC modality. Conversely, high frequencies (Channels 3 and 4) retain high weights, relying on AC features to compensate for the limited bandwidth of BC sensors. Notably, the mid-frequency band (Channel 2) shows the most dynamic adaptation, acting as a trade-off zone that actively balances BC stability and AC detail based on the instantaneous noise level.

\subsubsection{Convergence Behavior of the DBI Solver}
\label{subsubsec:demf_convergence}

\begin{figure}[ht]
    \centering
    \includegraphics[width=\linewidth]{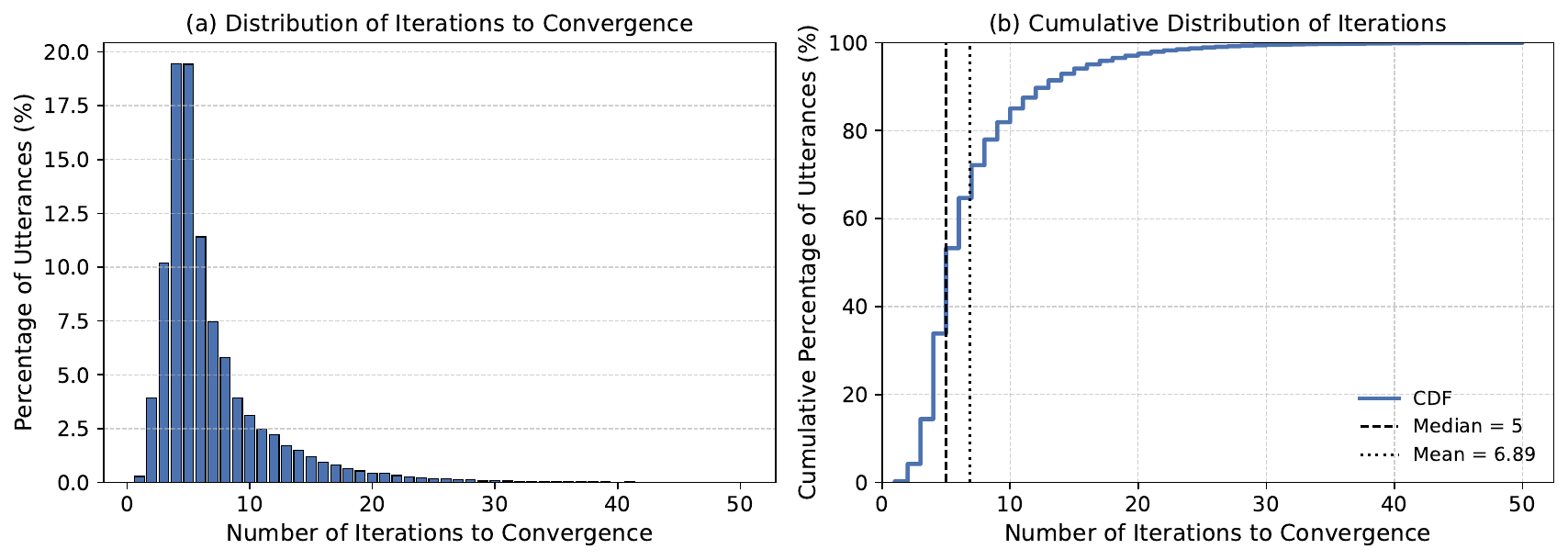}
    \caption{Iteration statistics of the DBI solver on the test set. 
    Panel (a) displays the distribution of the stopping iteration $k^{*}$ as a percentage of total utterances. 
    Panel (b) plots the CDF of $k^{*}$ with the median and mean values indicated.}
    \label{fig:iter_analysis}
\end{figure}

To quantify the practical inference cost, we analyze the distribution of the stopping iteration $k^{*}$ across the test set.
Here, $k^{*}$ represents the iteration index that minimizes the inter-iteration change $\delta_k$ within a bounded budget $K_{\max}$.
Fig.~\ref{fig:iter_analysis} summarizes the empirical distribution of $k^{*}$ alongside its Cumulative Distribution Function (CDF).

As illustrated in Fig.~\ref{fig:iter_analysis}(b), the DBI solver typically converges within a few refinement steps.
Specifically, the median stopping iteration is $k^{*}=5$, and the mean is $6.89$.
While certain challenging utterances exhibit a mild long-tail behavior, the iteration count remains far below the preset upper bound $K_{\max}=50$ in practice.
These results indicate that $K_{\max}$ serves primarily as a conservative safeguard.
Consequently, the fixed-point refinement terminates efficiently for the vast majority of utterances.
This ensures that the computational overhead during inference remains low and manageable.

\section{Conclusion} \label{sec:conclusion}
In this paper, we presented DBMIF, a deep balanced multimodal iterative fusion framework designed to robustly enhance speech by leveraging complementary AC and BC signals. The proposed architecture orchestrates a coarse-to-fine interaction strategy: it begins with an early DIAF module to adaptively weight modalities based on instantaneous reliability, follows with a multi-scale encoder–decoder equipped with CBGI for controlled bidirectional information exchange, and concludes with a parameter-shared DBI bottleneck that progressively refines the fused representation without expanding the model size. Experimental results confirm that the proposed framework consistently outperforms state-of-the-art unimodal and multimodal baselines across diverse noise environments.

Despite its generalization capability, the current non-causal architecture and high computational complexity pose challenges for real-time deployment. Furthermore, while DBMIF achieves marked improvements in extreme noise conditions, the intelligibility gap between the enhanced and clean speech remains noticeable, indicating the need for further innovation in ultra-low SNR regimes. Future work will focus on optimizing the model for causal, streaming inference on embedded platforms and exploring self-supervised pre-training to further boost the enhancement performance.

\bmhead{Acknowledgements}

This work was supported in part by the grants from the National Natural Science Foundation of China under Grant 62332019, the National Key Research and Development Program of China (2023YFF1203900, 2023YFF1203903), Sponsored by Beijing Nova Program (20240484513).

\bmhead{Data Availability}

The data used in this paper are all from public datasets.

\section*{Declarations}

\textbf{Competing interests} The authors declare that they have no known competing financial interests or personal relationships that could have appeared to influence the work reported in this paper.

\begin{appendices}




\end{appendices}


\bibliography{reference}

\end{document}